\documentclass[prb,twocolumn,showpacs,preprintnumbers,amsmath,amssymb]{revtex4}
\usepackage{graphicx}
\usepackage{dcolumn}
\usepackage{bm}

\begin{document}

\title{Density Functional Study of the Over-Doped Iron Chalcogenide:
TlFe$_{2}$Se$_{2}$ with ThCr$_{2}$Si$_{2}$ structure}

\author{Lijun Zhang}
\author{D.J. Singh}

\affiliation{Materials Science and Technology Division,
Oak Ridge National Laboratory, Oak Ridge, Tennessee 37831-6114}

\date{\today}

\begin{abstract}
We report density functional calculations of electronic structure and magnetic
properties of ternary iron chalcogenide TlFe$_{2}$Se$_{2}$, which occurs
in the ThCr$_{2}$Si$_{2}$ structure and discuss the results
in relation to the iron-based superconductors.
The ground state is antiferromagnetic with checkerboard
order and Fe moment $\sim$ 1.90 $\mu$B.
There is strong magnetoelastic coupling similar to the Fe-based
superconductors, reflected in a sensitivity of the Se position to
magnetism.
Tl is monovalent in this compound,
providing heavy electron-doping of 0.5 additional carriers per Fe
relative to the parent compounds of the Fe-based superconductors.
Other than the change in electron count, the electronic structure
is rather similar to those materials.
In particular, the Fermi surface is closely related to those
of the Fe-based superconductors, except that the electron cylinders are
larger, and the hole sections are suppressed. This removes
the tendency towards a spin density wave.
\end{abstract}


\maketitle

\section{introduction}

The finding of superconductivity in layered Fe oxy-pnictides with the ZrCuSiAs
structure
\cite{kamihara06,kamihara08}
has stimulated much interest leading to
the discovery of many additional compounds with
critical temperatures up to $\sim$ 56 K.\cite{wang-c,wu-g}
Superconductivity occurs quite generally in these compounds, and
is robust against off-site substitutions, and variations in the structure,
including doped fluoro-arsenides SrFeAsF,\cite{tegel,matsuishi}
as well as ThCr$_{2}$Si$_{2}$-type arsenides
(prototype BaFe$_{2}$As$_{2}$\cite{rotter-prl}),
Cu$_{2}$Sb-structure LiFeAs\cite{wang-xc} (NaFeAs\cite{parker}),
and PbO structure Fe(Se,Te).
\cite{hsu,mizuguchi}

All these materials have
square planar Fe$^{2+}$ layers,
tetrahedrally coordinated by anions and are near magnetism.
The undoped compounds generally show a spin density wave (SDW),
\cite{delacruz,rotter,torikachvili}
and superconductivity appears when this SDW is suppressed either
by pressure or doping.
This suggests an involvement of magnetism in
the pairing.\cite{mazin}
In addition there is accumulating evidence for
strong antiferromagnetic correlations in this
family in the normal metallic states, e.g. from the temperature
dependence of the susceptibility, which increases with $T$.
\cite{mcguire,klauss}

As mentioned, the binary chalcogenide $\alpha$-FeSe
is also superconducting when doped by off-stoichiometry.
\cite{hsu}
$T_c$ for this material reaches 27 K under pressure, \cite{mizuguchi}
while the electronic structure is very similar to the pnictide phases,
according to both density functional calculations \cite{subedi}
and photoemission.\cite{yoshida}
Among the chalcogenides, experiments indicated that
FeTe might have the strongest superconductivity based on the
increase in $T_c$ upon alloying FeSe with Te,
(increasing from 8 K to 15 K)\cite{yeh}
as well as theoretical considerations based on the strength of its SDW.
\cite{subedi}
However, FeTe shows an antiferromagnetic ground state
instead of superconductivity,\cite{fang,bao-w,li-sl}
which implies additional doping (or pressure) is still required to further
suppress magnetism and perhaps induce superconductivity.
Modest superconductivity (10 K) has already been observed when alloying
FeTe with S.
\cite{mizuguchi-s}
In this regard, it is important to note
that $\alpha$-FeSe and FeTe generally form
off stoichiometry with excess Fe partially filling interstitial sites
in the chalcogen layer.
This excess Fe acts as an electron dopant,
and suppresses SDW order by donating carriers to Fe-Se layers.
\cite{zhang}
In addition, the excess Fe,
carries local magnetic moments that may be expected to
interact with spin fluctuations associated with the Fe-Se layers
and also produce pair breaking.\cite{bao-w,zhang,fang-ex}
Therefore, it is of interest to explore alternate chemical doping
strategies for FeSe and FeTe, perhaps based on related compounds.

Here we report density functional studies of TlFe$_{2}$Se$_{2}$, which is
a ternary iron chalcogenide occurring in ThCr$_{2}$Si$_{2}$ structure.
Little is known about the physical properties of this compound,
except that it has been synthesized and that it has
a magnetic transition at $\sim$450 K, as observed in
M{\"o}ssbauer measurements. The magnetic state was characterized as
antiferromagnetic based on the lack of attraction of the magnetic
phase by a magnet. \cite{haggstrom}

\section{structure and methods}

\begin{figure}
\includegraphics[width=2.6in,height=3.4in]{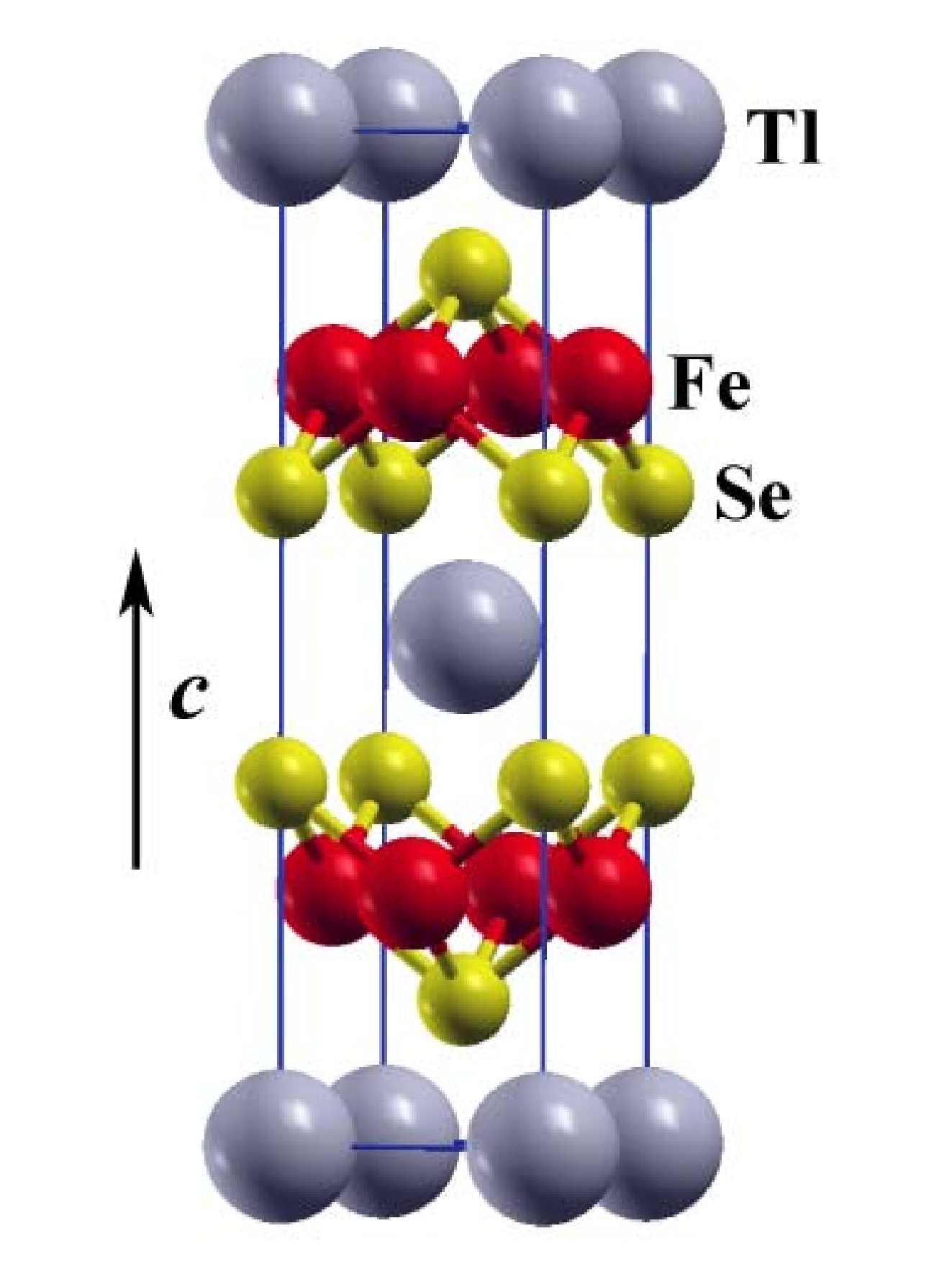}
\caption{(color online)
Crystal structure of ThCr$_{2}$Si$_{2}$-type TlFe$_{2}$Se$_{2}$.}
\label{stru}
\end{figure}

From a structural point of view,
TlFe$_{2}$Se$_{2}$ (Fig. \ref{stru}) is closely related to FeSe,
and in particular consists of practically identical Fe-Se layers,
consisting of edge-sharing FeSe$_{4}$ tetrahedra,
although these are now intercalated with Tl, altering the stacking
sequence.
The large size of Tl cations (even larger than
Ba in BaFe$_{2}$As$_{2}$) results
in a larger separation (7.0 {\AA}) between Fe-Se layers
than FeSe (5.5 {\AA}).
In our calculations, we used the
experimental values of tetragonal lattice constants,
$a$ = 3.89 {\AA} and $c$ = 14.00 {\AA},\cite{klepp}
while the internal coordinate, $z_{\rm Se}$ was relaxed by energy 
minimization
(the atomic coordinates in this structure are
Fe (4$d$) (0, 0.5, 0.25), Tl (2$a$) (0, 0, 0),
and Se (4$e$) (0, 0, $z_{\rm Se}$) )

The present calculations
were performed within the generalized gradient approximation (GGA) of
Perdew, Burke, and Ernzerhof (PBE),\cite{pbe} using
both the general potential linearized augmented plane-wave (LAPW)
\cite{singh-book,wien}
and projector augmented-wave (PAW) methods.\cite{kresse,vasp}
For the LAPW method, we employed LAPW spheres of radius 2.5$a_{\rm 0}$ for
Tl and 2.1$a_{\rm 0}$ for Fe and Se.
Well converged basis sets with the size determined
by $R_{\rm Fe}k_{\rm max}$=8.0 were used and semicore states
of 5$d$ for Tl, 3$p$ for Fe, and 3$d$, 4$s$ for Se were included using
local orbitals.
In the PAW calculations, a
kinetic energy cutoff of 350 eV and augmentation charge cutoff of 511 eV were
used.
A 16x16x16 grid was used for the body-centered tetragonal Brillouin zone
sampling in the self-consistent calculations,
while more dense grids were used for density of states
(DOS), Fermi surface, and especially magnetic calculations.
The electric-field gradient (EFG) calculations
were done with the all-electron LAPW method.
The formation energy for vacancies
was calculated with the PAW method using a 90 atom supercell, not
including magnetism.
The chemical potentials of bulk elemental Tl and Fe
was used as a reference. For $bcc$ Fe our calculated PAW spin
magnetic moment of 2.27 $\mu$B
is in reasonable agreement with experimental value 2.12 $\mu$B.
Consistency between the LAPW and PAW calculations was carefully cross-checked
and is indicated
for example by very small residual forces for the
relaxed structure, as well as very closely
coincident band structures and density of states with two methods.

\begin{figure}
\includegraphics[width=2.5in,height=3.4in,angle=270]{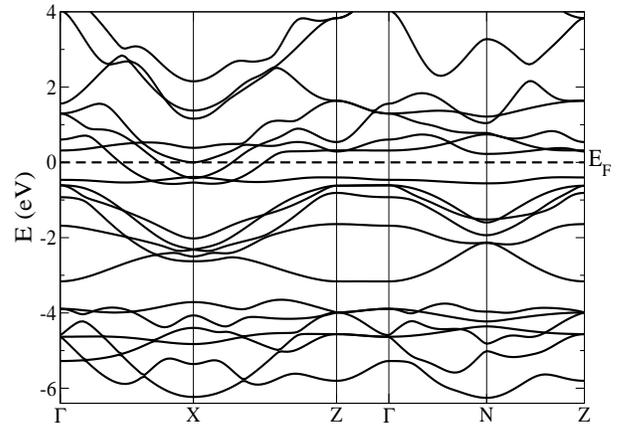}
\caption{(color online) Calculated electronic band structure of
TlFe$_{2}$Se$_{2}$ with the checkerboard antiferromagnetic order,
using the spin-polarized relaxed internal coordinate $z_{\rm Se}$ = 0.348.}
\label{bands_afm}
\end{figure}

\section{ground state: checkerboard antiferromagnetism}

We begin with the magnetic order of stoichiometric
TlFe$_{2}$Se$_{2}$, which is after all a known magnetic material.
The energetics were calculated with optimization of the
internal coordinate $z_{\rm Se}$ considering three
different long-range magnetic orders possibly existing within
Fe sheets:
\cite{mazin,mazin_prb}
ferromagnetism,
nearest-neighbor checkerboard antiferromagnetism,
and stripe antiferromagnetism (the pattern of the SDW in the
undoped Fe superconductors).
We found that the energy is always lowered in magnetic
configurations where the Fe layer stacking along the $c$ axis
in antiferromagnetic,
i.e. the interlayer Fe-Fe interaction is antiferromagnetic.
Relative to the non-spin-polarized case,
we found a ferromagnetic instability (-44 meV/Fe)
at the relaxed $z_{\rm Se}$ = 0.360,
with Fe moment 2.79 $\mu$B/Fe.
However
a stronger instability is found
for the checkerboard nearest neighbor antiferromagnetic order
(-78 meV/Fe, 1.90 $\mu$B/Fe) at relaxed $z_{\rm Se}$ = 0.348,
which is the ground state according to our calculations.
This seems to be consistent with the only available experimental
result, i.e. antiferromagnetism.\cite{haggstrom}
Obviously, further experiments will be helpful in confirming the
magnetic order in TlFe$_{2}$Se$_{2}$.
Interestingly, we did not find a magnetic instability at all
for the SDW-like ordering.
The sensitivity of the moment formation to the ordering pattern
shows that the magnetism is itinerant in nature.
In non-spin-polarized calculations, we obtain
a relaxed optimal internal Se coordinate of $z_{\rm Se}$=0.342,
lower than both the 
the experimental value 0.357 (with the difference of 0.2 {\AA})
and the value obtained in calculations including magnetism, although
with the lowest energy
checkerboard order, we still obtain a value significantly
lower than experiment.
In any case, considering also the difference between the
ferromagnetic and checkerboard values of $z_{\rm Se}$
we do find the substantial magnetoelastic coupling, similar to the Fe-based
superconducting phases.

\begin{figure}
\includegraphics[width=2.7in,height=3.4in,angle=270]{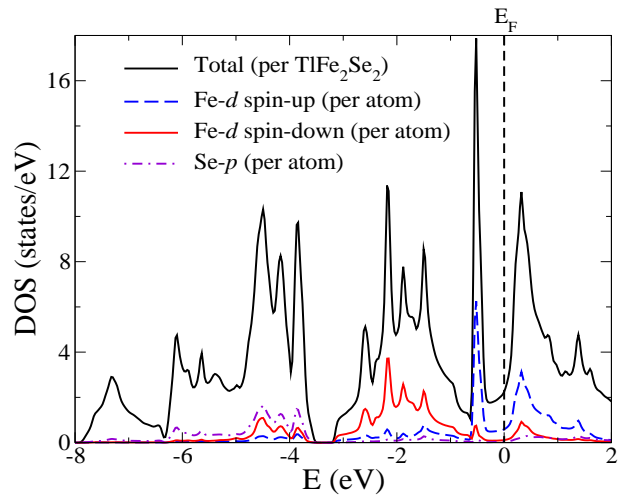}
\caption{(color online) Calculated total and projected electronic
DOS of TlFe$_{2}$Se$_{2}$
with the checkerboard antiferromagnetic order, using the LAPW method.
For Fe, the minority half-filled (spin-up) and
majority filled (spin-down) 3$d$ states are shown separately.}
\label{dos_afm}
\end{figure}

The calculated band structure and electronic
density of states (DOS) with the ground state checkerboard
antiferromagnetic order are shown in Figs. \ref{bands_afm} and \ref{dos_afm},
respectively.
This material is metallic
with two dispersive electron bands crossing the Fermi level
($E_{\rm F}$) around the $X$ point.
As may be seen, the Fermi energy occurs in a trough in the DOS
with a sharp peak $\sim$ 0.5 eV below $E_F$.
This peak arises from the very flat band below $E_{\rm F}$, as seen in 
Fig. \ref{bands_afm}.
At the Fermi energy,
$N(E_{\rm F})$=1.1 eV$^{-1}$ per Fe.
The Fe 3$d$ spin-down states are almost filled.
Integration of partial spin states up to $E_{\rm F}$ and normalization
with total Fe 3$d$ states give 4.3 electrons in the majority states
and 2.4 electrons in the minority states.
This indicates the valence state of Fe is lower than +2 here.
The Se $p$ states are mainly found below -3 eV relative to the
$E_{\rm F}$,
and are only modestly
hybridized with Fe $d$ states. The Tl 6$s$ states are occupied, while
the remaining Tl $6p$ states are above $E_F$, indicating that Tl
occurs as monovalent Tl$^{+}$. This is consistent with electron doping
of the the Fe-Se sheets, by 0.5 e / Fe, i.e. nominal Fe valence of Fe$^{1.5+}$.

To connect with future experiments,
we calculated the electric field gradient (EFG) of Fe and Se sites.
These are defined as the second derivative of the Coulomb
potential at the nuclear sites, and are
probed by nuclear magnetic resonance
(NMR) or nuclear quadrupole resonance (NQR) measurements.
For tetragonal site symmetry, only the $V_{\rm zz}$ component is
independent since the EFG tensor must be traceless.
The calculated $V_{\rm zz}$ with the ground state antiferromagnetic
order are -2.2x10$^{21}$ V/m$^{\rm 2}$
and -5.6x10$^{21}$ V/m$^{\rm 2}$ for Fe and Se, respectively.
As noted below these values are very sensitive to the value
of the internal coordinate, and therefore EFG measurements
may be a very useful probe of magnetoelastic coupling in these
materials.
In this regard, it should be noted that Mukuda and co-workers 
observed a relationship between the As quadrupole frequency
and $T_c$ in oxy-arsenides. \cite{mukuda}

\section{Relationship with the iron superconductors}

\begin{figure}
\includegraphics[width=2.5in,height=3.4in,angle=270]{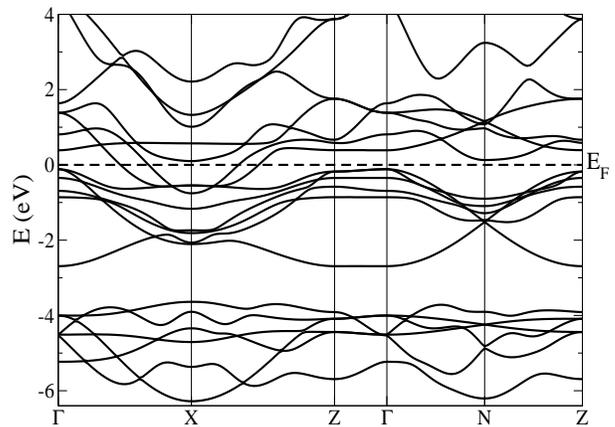}
\caption{(color online) Calculated non-spin-polarized band structure of TlFe$_{2}$Se$_{2}$ using the relaxed $z_{\rm Se}$ = 0.342.}
\label{bands}
\end{figure}

\begin{figure}
\includegraphics[width=2.7in,height=3.4in,angle=270]{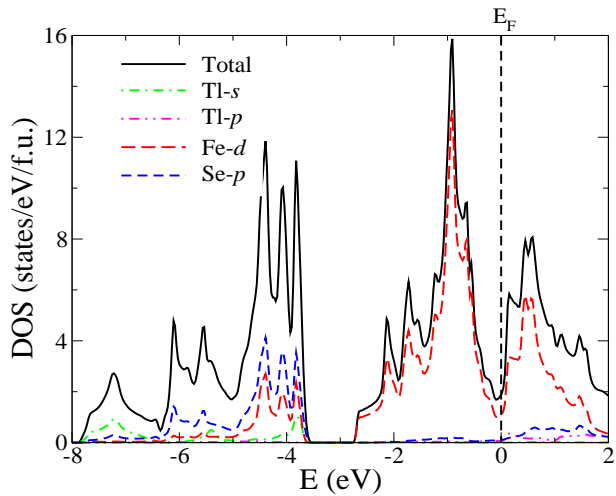}
\caption{(color online)
Calculated non-spin-polarized total and projected electronic DOS
of TlFe$_{2}$Se$_{2}$.
The values are on a per formula unit basis.
Note that projections are onto the LAPW spheres,
thus the absolute contributions of the Tl-$s$ and Se-$p$ states are 
underestimated owing to their more extended orbitals which
extend significantly into the interstitial region.}
\label{dos}
\end{figure}

\begin{figure}
\includegraphics[width=3.4in]{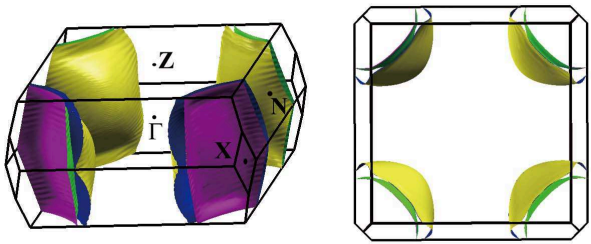}
\caption{(color online) Calculated non-spin-polarized Fermi surface of
TlFe$_{2}$Se$_{2}$.
The right panels are top views along $Z-\Gamma$ direction.
Here Fermi surfaces were mapped in the body-centered tetragonal Brillouin zone,
and thus $X$ point corresponds to the zone corner $M$ point in the tetragonal
Brillouin zone.}
\label{fermi}
\end{figure}

To clearly show the relation of TlFe$_{2}$Se$_{2}$ with the Fe-based
superconductors, we performed non-spin-polarized calculations of
the electronic
structure.
The band structure, DOS, and Fermi surface are shown in
Figs. \ref{bands}, \ref{dos} and \ref{fermi}, respectively.
The main features of band structure show close similarity with that of
FeSe\cite{subedi}
and the Fe-based arsenides.
\cite{singh-du,bafe2as2}
In particular, the states near $E_{\rm F}$ are dominated by Fe $d$ states
(mainly derived from $xz$, $yz$ orbitals)
forming a pseudogap at the electron count of 6
(rather than 4 in the tetrahedral crystal field scheme), with
some mixture with Se $p$ states.
Tl occurs as Tl$^{+}$, as mentioned, with
the Tl $s$ states in the energy range
$\sim$ -8 eV to -4 eV and the $p$ derived valence states are
above $E_{\rm F}$.
The resulting doping relative to FeSe (0.5 e/Fe)
pushes $E_{\rm F}$ up
from the steep lower edge of the pseudogap for FeSe\cite{subedi}
to the bottom,
leading to significantly decreased $N(E_{\rm F})$ of 0.92 eV$^{-1}$ per Fe.
As a result, the hole Fermi surfaces at the zone center in other
Fe-based materials have completely disappeared and the electron
sections at zone corner greatly expand, as shown in Fig. \ref{fermi}.
This destroys the strong nesting between two-dimensional electron
and hole Fermi surfaces in more lightly doped Fe-based materials
and thus the SDW antiferromagnetic instability is suppressed.

As in the other Fe-based superconducting materials,
\cite{mazin_prb}
the non-spin-polarized electronic structure of TlFe$_{2}$Se$_{2}$ also shows
great sensitivity to the Se height.
For example, if the experimental $z_{\rm Se}$ = 0.357 was used for
calculations,
$N(E_{\rm F})$ increases to
2.0 eV$^{-1}$ per Fe though $E_{\rm F}$ still lies in the bottom of pseudogap,
and furthermore very small cylindrical
hole Fermi surfaces appear at the zone center.
This increase in $N(E_F)$ with $z_{\rm Se}$
provides an explanation for the fact that, although not the
ground state, an
itinerant ferromagnetic state is lower in energy than the non-magnetic
state
even though the non-spin-polarized $N(E_F)$ (with the non-spin-polarized
$z_{\rm Se}$)
is well below the Stoner criterion for itinerant magnetism.

The calculated $V_{\rm zz}$ of Fe and Se are
1.0x10$^{21}$ V/m$^{\rm 2}$ and -4.4x10$^{21}$ V/m$^{\rm 2}$
with experimental $z_{\rm Se}$ and -0.4x10$^{21}$ V/m$^{\rm 2}$, and
-5.2x10$^{21}$ V/m$^{\rm 2}$
with relaxed $z_{\rm Se}$, respectively.
This illustrates
the strong sensitivity of the EFG to the Se position, as mentioned above.
Strong sensitivity of the EFG
was also found for As in LaFeAsO
by density functional calculations.\cite{Grafe}

\section{summary and discussion}

To summarize, based on density functional calculations,
we study magnetic properties, electronic structure and relation
to superconductivity for TlFe$_{2}$Se$_{2}$
occurring as the ThCr$_{2}$Si$_{2}$ structure.
We find the nearest-neighbor checkerboard antiferromagnetism is the
ground state, consistent with the available experimental data.
The
non-spin-polarized electronic structure of TlFe$_{2}$Se$_{2}$ shows close
similarity with the Fe-based superconductors.
Relative to those materials, Tl${^+}$ is an electron dopant,
donating 0.5 additional carrier per Fe relative to Fe-Se layers.
This over-doping significantly enlarges the electron sections of Fermi surface
at zone corner and eliminates hole sections at zone center, and thereby
completely destroys the Fermi nesting and thus suppresses the
SDW antiferromagnetic instability.
We note that the compound crystallizes in the ThCr$_{2}$Si$_{2}$ structure,
with a larger interlayer separation than FeSe.
This might be favorable to superconductivity by empirically considering
that the highest $T_{c}$ so far
is realized in the ZrCuSiAs-type compounds, i.e. the family with the
largest interlayer space.\cite{wang-c,wu-g} 
In any case, TlFe$_2$Se$_2$ is not reported to have
excess Fe in interstitial sites whose moments may
interact with superconducting Fe-Se planes and cause pair breaking.
However, some Fe vacancies might form as the intrinsic defects in this
system.\cite{haggstrom,sabrowsky}
One natural way forward is via Tl deficiency, which would reduce the
over-doping.

In order to
evaluate this possibility, we calculated the formation energy of a Tl vacancy,
using a supercell as described above.
The calculated energy 
of -0.14 eV, indicates metastability of the
compound at low temperature, similar to PbZrO$_3$, \cite{navrotsky,kagimura}
and that abundant Tl vacancies
are very prone to form in this material.
Thus, it may well be possible to produce Tl deficient material by
modifying the growth conditions.
\cite{heat}
Owing to absence of excess Fe that cause pair breaking and suppress
superconductivity in FeSe, if appropriate Tl deficiency introduced,
this material (Tl$_{x}$Fe$_{2}$Se$_{2}$) might be a good candidate for
higher $T_{c}$ in chalcogenide family. This structure type may
also provide an avenue for producing superconductivity in Fe-Te compounds,
e.g. Tl$_{x}$Fe$_2$Te$_2$.

Even in the over-doped regime represented by TlFe$_2$Se$_2$,
where the Fermi level has dropped into the bottom of the pseudogap,
resulting in relatively low $N(E_{\rm F}$),
we still found competing itinerant
ferromagnetic and checkerboard antiferromagnetic states.
In this regard, recent NMR experiments have shown evidence
for pseudogap behavior in electron doped compounds (note this is the NMR
pseudogap, not the pseudogap in the DOS),
perhaps related to that in cuprates. \cite{warren,alloul}
However, the behavior is different. In cuprates the pseudogap
appears to be closely associated with the magnetic ordering of the
undoped phases, and is seen most strongly in the underdoped
regime. In contrast, the Fe-based superconductors show the
pseudogap in the electron over-doped regime and not in the
underdoped regime.
\cite{nakai-2,ning}
However, while Fe$^{2+}$ is a common valence for Fe, Fe$^{1+}$ is
not and perhaps for this reason heavily over-doping beyond the
superconducting regime while maintaining high sample
quality has been difficult.
Within an itinerant picture one may expect competition of
SDW order and other magnetic (i.e. checkerboard) order
in the Fe superconductors, which gives way to checkerboard order
when the hole Fermi surface, and therefore the nesting is destroyed
by over-doping. Within such a scenario, and
considering the experimental results, the pseudogap may be associated
with incipient checkerboard order. In this regard, the chemical
stability of TlFe$_2$Se$_2$ may provide a very useful window into
the over-doped regime, especially if it can be chemically doped with
holes, e.g. by Tl deficiency, as discussed above.

In fact, such such studies might be particularly illuminating considering
the phase diagram that is suggested by the present results in conjunction
with the known phase diagrams of the electron doped superconductors,
first principles results for them and NMR data.
In particular, without doping the materials show a spin density wave, though
with strongly reduced moments compared to standard density functional
calculations. Such calculations do not include renormalization due to spin
fluctuations -- an effect that is small in most magnetic materials, such
as Fe metal, but is apparently large in these materials. Electron (or hole)
doping destroys the SDW, presumably by reducing the nesting between the
hole and electron Fermi surfaces, in favor of a metallic state with 
evidence for strong spin fluctuations and superconductivity possibly
related to spin fluctuations connected with the SDW.
In this regime, there also appears to be a strong renormalization
of magnetism, based on comparison of density functional results and
experiment. \cite{mazin_prb,singh-physica}
In general such
renormalized states occur near quantum critical points or in situations
where there is a competition between different magnetic orders.
Considering the large composition range over which this behavior is seen,
competition with another magnetic state may be crucial. The experimental
observation that TlFe$_2$Se$_2$ is magnetic, and the likely checkerboard
order suggests that the competing state in the superconducting phases
might be the nearest neighbor antiferromagnetic state. This state, although
itinerant, is not driven by Fermi surface nesting, and therefore nearness to 
it would
be expected to yield spin fluctuations over a broader region of the
Brillouin zone than nearness to a nesting driven SDW. While this would
make them more effective in renormalizing the SDW instability due to the
larger phase space for competing fluctuations, it would also make them
harder to observe by inelastic neutron scattering as compared to the nesting
related fluctuations. Nonetheless, it would be of great interest first
of all to confirm experimentally that the ground state of TlFe$_2$Se$_2$
is the checkerboard state, and secondly to study the overdoped regime, perhaps
using this chemistry, to relate
spin fluctuations from neutron scattering and the NMR pseudogap.
Furthermore, it will be of considerable interest to study the interplay
between superconductivity and the onset of checkerboard order, if this
in fact is established at high doping. While the SDW and superconductivity
mediated by nesting related spin fluctuations would compete for the
same electrons at the Fermi surface, this is not necessarily the
case for the checkerboard order, and so some difference may be
anticipated from the behavior at the underdoped antiferromagnetic (SDW) -
superconducting boundary,
perhaps including coexistence of the two orders over
some composition range.

\section{acknowledgments}

We are grateful for helpful discussions with M.H. Du, A. Subedi, and D. Mandrus.
Some figures were produced with the XCRYSDEN
program.\cite{kokalj}
This work was supported by the Department of Energy,
Division of Materials Sciences and Engineering.

\bibliography{TlFe2Se2}

\begin{thebibliography}{48}
\bibitem[{hea()}]{heat}
\bibinfo{note}{Since the negative formation energy of Tl vacancies corresponds
  to metastability of the compound, we also checked the heat of formation for
  TlFe$_{2}$Se$_{2}$ in terms of elements. The obtained value is -1.02 eV/f.u.
  (98 kJ/mol/f.u.), which indicates it is at least stable against decomposition
  into elements.}
\end{thebibliography}
\end{document}